\documentclass[aps, prl, reprint, showpacs, showkeys, twocolumn, superscriptaddress, notitlepage]{revtex4-1}

\usepackage{graphicx} 
\usepackage[english]{babel}
\usepackage{braket}
\usepackage{gensymb}
\usepackage{color}
\usepackage{amssymb}
\usepackage{times}
\usepackage{textcomp} 
\usepackage{amsmath}

\usepackage[colorlinks=true,citecolor=blue,linkcolor=magenta]{hyperref}
\hypersetup{
pdfauthor = {P. Maletinsky},
colorlinks = true, linkcolor = blue, urlcolor=blue, bookmarksnumbered =  true}

{
\begin{document}
\title{Hybrid continuous dynamical decoupling: a photon-phonon doubly dressed spin}

\author{Jean Teissier, Arne Barfuss, Patrick Maletinsky}
\affiliation{Quantum Sensing Group, Department of Physics, University of Basel, Klingelbergstrasse 82, CH-4056 Basel, Switzerland}

\date{\today}

\begin{abstract}

We study the parametric interaction between a single Nitrogen-Vacancy electronic spin and a diamond mechanical resonator in which the spin is embedded.  
Coupling between spin and oscillator is achieved by crystal strain, which is generated upon actuation of the oscillator and which parametrically modulates the spins' energy splitting. 
Under coherent microwave driving of the spin, this parametric drive leads to a locking of the spin Rabi frequency to the oscillator mode in the megahertz range. 
Both the Rabi oscillation decay time and the inhomogeneous spin dephasing time increase by two orders of magnitude under this spin-locking condition. We present routes to prolong the dephasing times even further, potentially to the relaxation time limit. 
The remarkable coherence protection that our hybrid spin-oscillator system offers is reminiscent of recently proposed concatenated continuous dynamical decoupling schemes and results from our robust, drift-free strain-coupling mechanism and the narrow linewidth of the high-quality diamond mechanical oscillator employed. 
Our findings suggest feasible applications in quantum information processing and sensing.
\end{abstract}

\maketitle

Solid-state spins rank amongst the most promising sources for quantum information processing and sensing, due to their ease of use and the in-principle scalability they offer.
Exploiting their quantum nature for computation or sensing requires quantum coherence to be preserved for a time long compared to the speed of their coherent manipulation. 
The fundamental limit of these coherence times is set by the spin relaxation rate $T_1$\,\cite{Slichter2013,Myers2016} --- in some cases exceeding seconds\,\cite{Amasha2008} --- while spin manipulation rates in the Gigahertz range have recently been demonstrated\,\cite{Fuchs2009}.
Despite these interesting prospects, reaching relaxation-limited coherence times for solid state spins remains highly challenging due to additional noise sources in the spins' environment and driving fields, which significantly reduce the number of coherent spin-manipulation steps that can be experimentally achieved.

\begin{figure}[th!]
\includegraphics[scale=1]{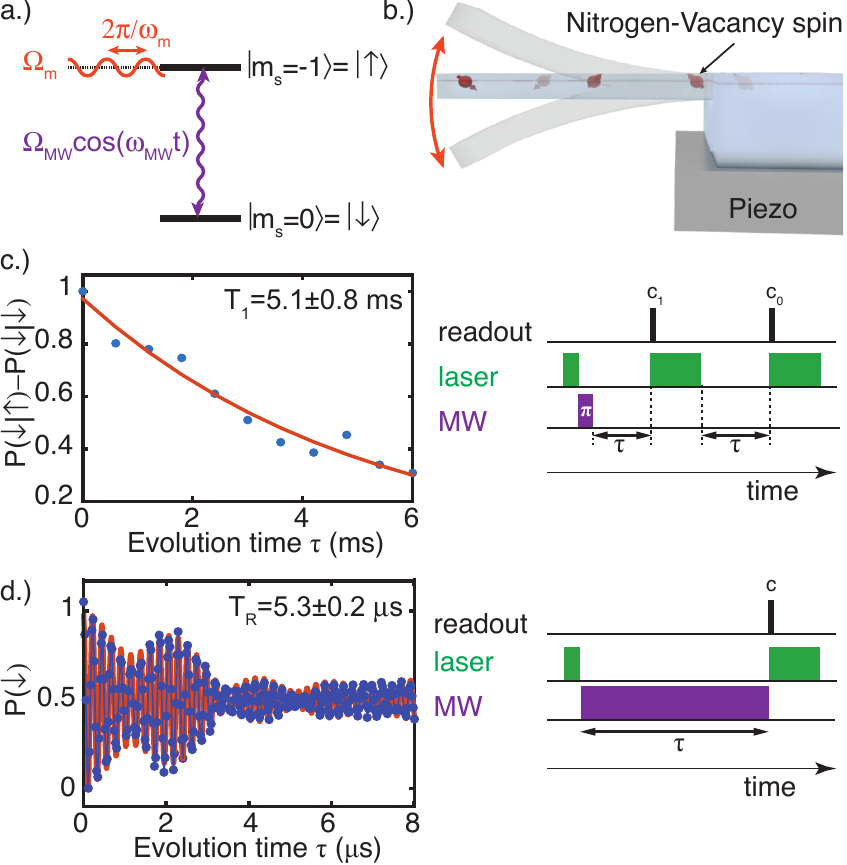} 
\caption{\label{FigOverview} 
a.) Relevant spin levels involved in our experiment. The resonant driving field and parametric strain-drive are indicated by purple and orange wavy lines, respectively.
b.) Schematic of the experimental device. A Nitrogen-Vacancy (NV) spin is embedded in a diamond cantilever, which is actuated using a nearby piezoelectric transducer to generate time-varying strain for parametric driving.
c.) Spin-relaxation and 
 d.) coherent Rabi oscillations of the NV spin, with corresponding measurement pulse sequences. 
Exponential and Gaussian fits (orange) to the data (blue) yield decay times $T_1=5.1\pm0.8~$ms and T$_R=5.3\pm 0.2~\mu$s for population and oscillation decay-times, respectively. 
}
\end{figure}

Various approaches to enhance spin coherence times towards the relaxation time limit have been put forward.
Most notably, dynamical decoupling --- a method of filtering out environmental noise through either pulsed\,\cite{Viola1999,Viola1998} or continuous\,\cite{Fanchini2007,Golter2014a,Barfuss2015,MacQuarrie2015} driving of the spin --- can isolate the spins from the low-frequency environmental fluctuations responsible for dephasing.
In this regard, pulsed schemes have proven especially effective\,\cite{BarGill2013}, robust to pulse errors\,\cite{DeLange2010} and even allow for decoherence protected quantum gates\,\cite{VanderSar2012}. However, they come at the cost of increased experimental complexity and potentially harmful, high intensity driving pulses. 
Decoupling schemes relying on continuous driving, on the other hand, are experimentally simpler\,\cite{Fanchini2007,Golter2014a}, do not suffer from pulsing errors, and can be readily combined with quantum gate operations\,\cite{Timoney2011,Xu2012}. 
Unfortunately, the effectiveness of these schemes is limited by the spin's high sensitivity to the ubiquitous low-frequency fluctuations of the driving field. 
As recently shown\,\cite{Khodjasteh2005,Cai2012a}, applying higher order drivings fields, with each additional field decoupling the spin from the driving field fluctuations of the preceding driving field, would protect the spin even further.
In principle, this procedure can be iterated ad infinitum and may then yield relaxation limited coherence times\,\cite{Cai2012a}. 
The application of many consecutive decoupling fields, however, exposes the spin to significant driving field powers and sets intrinsic constraints to the speed at which the final, decoherence protected spin states can be coherently manipulated\,\cite{Cai2012a}. 
New approaches to continuous dynamical decoupling are therefore required to yield fully robust spin systems which are of practical use to quantum information processing and sensing.

In this work, we experimentally demonstrate a novel and efficient approach to continuous dynamical decoupling, through the parametric interaction of a single electronic spin with a mechanical resonator. We employ a coherent microwave drive for first order decoupling and use the spin-oscillator interaction to decouple the spin from amplitude fluctuations in the microwave field. 
This concatenated, hybrid continuous dynamical decoupling (HCDD) builds on two key advances over past approaches\,\cite{Cai2012a}: 1. ) Second order decoupling is achieved by a parametric drive along the quantisation axis of the undriven spin. The second order driving field is thus orthogonal to the first order drive, irrespective to the phase between these two fields, in contrast to the conventional dynamical decoupling by concatenated driving, where phase-locking between the driving field is required to ensure the necessary orthogonality.  
2.) The second order decoupling field is transduced to the spin through a mechanical oscillator, whose resonant behaviour effectively low-pass filters amplitude noise and yields a highly stable, second order driving field amplitude. 
A concatenation of only two driving fields thereby yields a coherence time nearly two orders of magnitude longer than that of an undriven spin, while also maintaining a final dressed state splitting close to one MHz.
By using mechanical oscillators with even higher quality-factors, our scheme should allow us to prolong coherence times even further and ultimately reach the limit imposed by energy relaxation. 

Our experiments were performed on individual, negatively charged Nitrogen-Vacancy (NV) defect centers embedded in singly-clamped cantilever diamond mechanical oscillators (Fig.\,\ref{FigOverview}b). 
The cantilevers exhibited typical fundamental mode-frequencies $\omega_{\rm m}$ in the MHz range ($\omega_{\rm m}=2\pi\times5.81$~MHz for the cantilever studied here) and were fabricated from ultra-pure, single-crystal, synthetic diamond using top-down nanofabrication  described elsewhere\,\cite{Maletinsky2012,Ovartchaiyapong2012}.
NV centers were created at densities $<1~\mu$m$^{-2}$ by $^{14}$N ion implantation and subsequent high-temperature annealing\,\cite{Chu2014}. 
We used a homebuilt confocal microscope to study spin-dynamics of an NV center located at the base of the cantilever, where strain-fields for parametric driving are maximised\,\cite{Teissier2014,Ovartchaiyapong2014}. All our experiments were performed under ambient conditions.

The NV center orbital ground state is a spin-triplet, with $\hat{S_z}$ eigenstates $\ket{-1},\ket{0},\ket{+1}$, where $\hat{S_z}$ is the angular momentum operator along the NV binding axis $z$. The magnetic sublevels $\ket{\pm1}$ are split from $\ket{0}$ by a zero-field splitting $D_0=2.87$~GHz and can be further split in energy by a magnetic field $B_z$ along $z$\,\cite{Doherty2013}. Optical spin-readout and initialisation into $\ket{0}$ is readily achieved by green optical illumination and detection of red NV fluorescence, while the spin can be coherently driven by applying resonant microwave magnetic fields, $B_\perp^{\rm AC}$, transverse to $z$\,\cite{Jelezko2004a,Gruber1997}.
In this work, we consider the dynamics of the effective two-level system formed by $\ket{0}=:\ket{\downarrow}$ and $\ket{-1}=:\ket{\uparrow}$, while the state $\ket{1}$ is split off in energy by a static magnetic field $B_z=10.7~$G 
and ignored in the following.  
In addition to near-resonant microwave driving with transverse magnetic fields, we employ parametric driving by time-varying (AC) strain fields along $z$ generated by the cantilever (Fig.\,\ref{FigOverview}a). 
The effective Hamiltonian for the two-level system spanned by $\{\ket{\downarrow},\ket{\uparrow}\}$ then reads
\begin{equation}
H/h=(D_0-\gamma_{NV}B_z+d_\parallel\Pi_\parallel^{\rm AC})\sigma_z+\gamma_{NV}B_\perp^{\rm AC}\sigma_x,
\label{EqHamiltonian} 
\end{equation}
with  $\sigma_i$ the Pauli matrices along direction $i\in\{x,z\}$, $\gamma_{NV}=2.8~$MHz/G the NV gyromagnetic ratio and ${\rm B}_{\perp}^{\rm AC}=(\Omega_{\rm MW}/2\pi\gamma_{NV})\cos\left(\omega_{\rm MW}t\right)$ the microwave driving field with frequency $\omega_{\rm MW}$ and amplitude (Rabi frequency) $\Omega_{\rm MW}$.
Parametric driving of the spin is achieved by on-axis, AC strain $\Pi_\parallel^{\rm AC}=\Pi_\parallel^{0}\cos\left(\omega_m {\rm t}\right)$ with peak amplitude $\Pi_\parallel^{0}$, induced by the cantilever oscillation. Such parallel strain couples to the NV center with strength $d_{\parallel}\sim5.5~$GHz/strain\,\cite{Teissier2014,Ovartchaiyapong2014} while transverse, AC strain can be neglected in our analysis as it was weak for the present NV and off-resonant from any involved spin transition\,\cite{Barfuss2015}.

To provide a baseline for our subsequent measurements, we first determined the relevant NV spin relaxation times in the absence of mechanical driving. 
The NV population decay time $T_1$
was determined using the experimental pulse sequence illustrated in Fig.\,\ref{FigOverview}c. 
Following initialisation in $\ket{\downarrow}$ ($\ket{\uparrow}$), we determine the difference $\Delta P=P(\downarrow|\uparrow)$-$P(\downarrow|\downarrow)$, as function of the variable delay $\tau$, where $P(i|j)$ is the population in state $\ket{i}$, after initialization in state $\ket{j}$ ($i,j\in\{\downarrow,\uparrow\}$). 
We obtain $\Delta P$ directly from the transient NV fluorescence photons $c_1$ and $c_0$, as defined in Fig.\,\ref{FigOverview}c, as $\Delta P(\tau)=(c_1(\tau)-c_0(\tau))/(c_1(0)-c_0(0))$.
Therefore, $\Delta P$ yields a measure for the spin population decay, from which we determine T$_1=5.1\pm0.8$~ms through an exponential fit (Fig.\,\ref{FigOverview}c).
Similarly, we determined the decay-time $T_{\rm R}$ of the spin's Rabi oscillations by pulsed, coherent driving of the $\ket{\uparrow}\leftrightarrow\ket{\downarrow}$ transition with a resonant microwave magnetic field of variable duration $\tau$ (Fig.\,\ref{FigOverview}d). 
The observed Rabi oscillations show a pronounced beating pattern that results from the $\sim2.18~$MHz hyperfine-splitting 
between the NV electronic spin and the Nitrogen's $^{14}$N nuclear spin ($I_{^{14}N}$=1)\,\cite{He1993}. 
Our data is well fit by $\sum_{j\in{-1,0,1}}{\rm A}_i \cos(\Omega_{\rm eff}^j \tau)e^{-(\tau/T_{\rm R})^2}$ (Fig.\,\ref{FigOverview}d, orange), where $\Omega_{\rm eff}^j=\sqrt{\Omega_{\rm MW}^2+\delta_j^2}$  
are the effective Rabi frequencies and $\delta_j$ the detunings between the microwave drive and each of the three hyperfine transitions. 
From the fit, we find $\Omega_{\rm MW}=2\pi \times5.81$~MHz and $T_{\rm R}=5.3\pm0.2~\mu$s, which is three orders of magnitude shorter than the relaxation-limit set by $T_1$\,\cite{Slichter2013}. 

\begin{figure}[t!]
\includegraphics[scale=1]{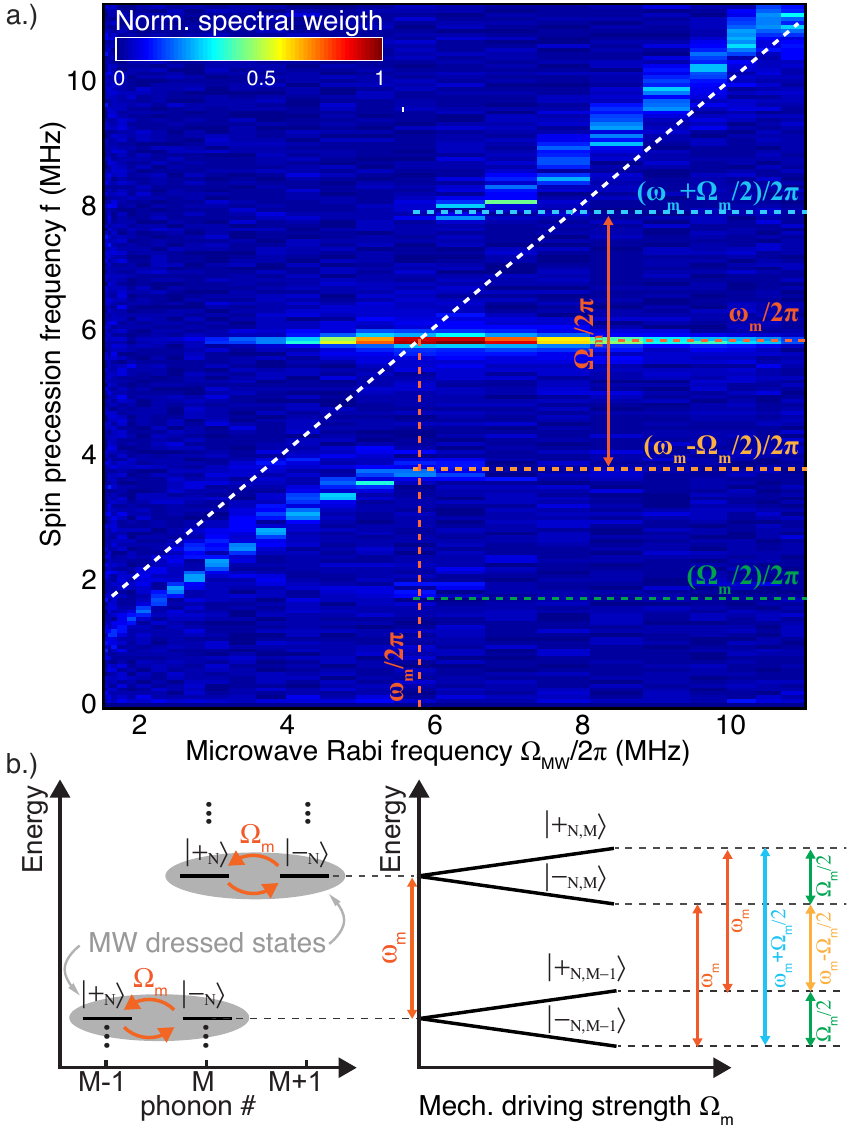} 
\caption{\label{FigMollow} 
a.) Spin precession spectrum under combined microwave drive and parametric mechanical strain-driving, as a function of microwave Rabi frequency $\Omega_{\rm MW}$. The spectra were obtained by Fourier-transforming the experimentally acquired Rabi oscillation data for each $\Omega_{\rm MW}$. The white dashed line follows $f=\Omega_{\rm MW}/2\pi$, i.e. the precession frequency expected for pure microwave driving within the rotating wave approximation. Coloured dashed lines indicate the characteristic frequencies occurring in the spin precession spectrum of the doubly-dressed spin. 
b.) Eigenenergies ($\hbar=1$) of the doubly dressed spin under resonant microwave driving and resonant parametric driving  ($\Omega_{\rm MW}=\omega_m$). 
Left: Energies of microwave dressed spin states $\ket{\pm_N}$ as a function of phonon number in the cantilever. For resonant driving, $\ket{+_N,M-1}$ and $\ket{-_N,M}$ are degenerate and coupled by the parametric drive with amplitude $\Omega_m$ (orange arrow), leading to the second order dressed states $\ket{\pm_{N,M}}$ (see text).  
Right: Eigenenergies of $\ket{\pm_{N,M}}$ and $\ket{\pm_{N,M-1}}$ as a function of $\Omega_m$. Coloured arrows indicate allowed dressed-state transitions, which are reflected by corresponding features in the spin precession spectrum shown in a.).
}
\end{figure}

The Gaussian decay-envelope of our Rabi oscillations suggests that slowly fluctuating noise sources are responsible for the excess dephasing we observed\,\cite{Kubo2012}. While both, $\Omega_{\rm MW}$ and $\delta_j$ may fluctuate, in our experiment, where $\Omega_{\rm MW}\gtrsim2\delta_j$, only the former contributes to first order to dephasing.
Indeed, we were able to quantitatively reproduce the observed decay envelope (orange line in Fig.\,\ref{FigOverview}d) by numerically averaging over Rabi oscillations with random microwave amplitude noise of relative amplitude $6\times 10^{-3}$ --- a typical value for the commercial microwave sources we employ.

In the presence of continuous, resonant microwave driving, the eigenstates of the driven spin system are the ``dressed states'' $\ket{\pm_N}:=\left(\ket{\downarrow,N}\pm\ket{\uparrow,N-1}\right)/\sqrt2$\,\cite{Cohen-Tannoudji1992,Cai2012a}, with energy difference $\hbar\Omega_{\rm MW}$ and $N$ the number of microwave photons dressing the spin (i.e. the mean photon number in the coherent microwave field which drives the spin). 
These new basis states form a potential resource for quantum information processing\,\cite{Timoney2011} or sensing\,\cite{Cai2012a}. The Rabi decay time $T_{\rm R}$, which can be interpreted as the 
dressed state relaxation time\,\cite{Rohr2014phd},  is then a key figure of merit for such applications. 
To further prolong $T_R$, we decouple $\ket{\pm_N}$ from fluctuations in $\Omega_{\rm MW}$ by applying an additional driving field, which near-resonantly and coherently drives the $\ket{+_N}\leftrightarrow\ket{-_N}$ transition (Fig.\,\ref{FigMollow}b) and consequently lead to higher-order dressed states --- the principle underlying dynamical decoupling by concatenated continuous driving\,\cite{Cai2012a}.

We achieve second order dressing and the associated dynamical decoupling by driving the $\ket{+_N}\leftrightarrow\ket{-_N}$ transition using the time-varying, longitudinal strain field generated by the diamond cantilever in which the NV spin is embedded. Such driving is enabled by the coupling term $d_\parallel\Pi_\parallel^0\sigma_z$ (c.f. eq.\,(\ref{EqHamiltonian})), which drives the desired transition at a rate $\Omega_m=d_\parallel\Pi_\parallel^0\bra{+_N}\sigma_z\ket{-_N}\neq0$. Resonance of that second drive tone with the dressed state energy splitting can be achieved by adjusting $\Omega_{\rm MW}$ to fulfil $\Omega_{\rm MW}\approx\omega_m$, while $\omega_m$ is fixed and given by the cantilever geometry.

To demonstrate second order dressing and the subsequent coherence protection, we performed resonant Rabi oscillation measurements up to an evolution time $\tau_{\rm max}=16~\mu$s at variable $\Omega_{\rm MW}$ and in the presence of a parametric, cantilever-induced strain-drive of fixed amplitude $\Omega_m\approx2\pi\times4.1~$MHz. Figure\,\ref{FigMollow}a shows the Fourier transformation of each experimental Rabi oscillation as a function of $\Omega_{\rm MW}$.
For $\Omega_{\rm MW}$ far from the dressed-state transition energy (i.e. for $\left|\Omega_{\rm MW}-\omega_m\right|\gtrsim\Omega_m$), the spin precession dynamics are dominated by a single peak at frequency $\Omega_{\rm MW}/2\pi$, as expected for conventional, coherent spin driving (white dashed line in Fig.\,\ref{FigMollow}a). 
The weak, additional spectral features visible in this regime stem from the two additional, hyperfine-split NV spin transitions, which are weakly, off-resonantly driven. 
For $\Omega_{\rm MW}\approx\omega_m=2\pi\times5.81~$MHz, however, we observe a spectrum that shares striking similarities with the well-known Mollow-triplet in quantum electrodynamics\,\cite{Rohr2014,Pigeau2015}: the measured coherent spin oscillations peak at a single frequency $\omega_m/2\pi$, irrespective of the exact value of $\Omega_{\rm MW}$, with only two weak side bands appearing at $(\omega_m\pm\Omega_m/2)/2\pi$. 
Such spin-oscillator frequency-locking, induced by parametric strain-driving of a bulk NV center in a mechanical oscillator, was previously observed for NV centers in diamond nanocrystals parametrically driven by the magnetic fields from a nearby antenna\,\cite{Rohr2014} or by the mechanical motion of a spin in a strong magnetic field gradient\,\cite{Pigeau2015}.

\begin{figure*}[t!]
\includegraphics[width=18cm]{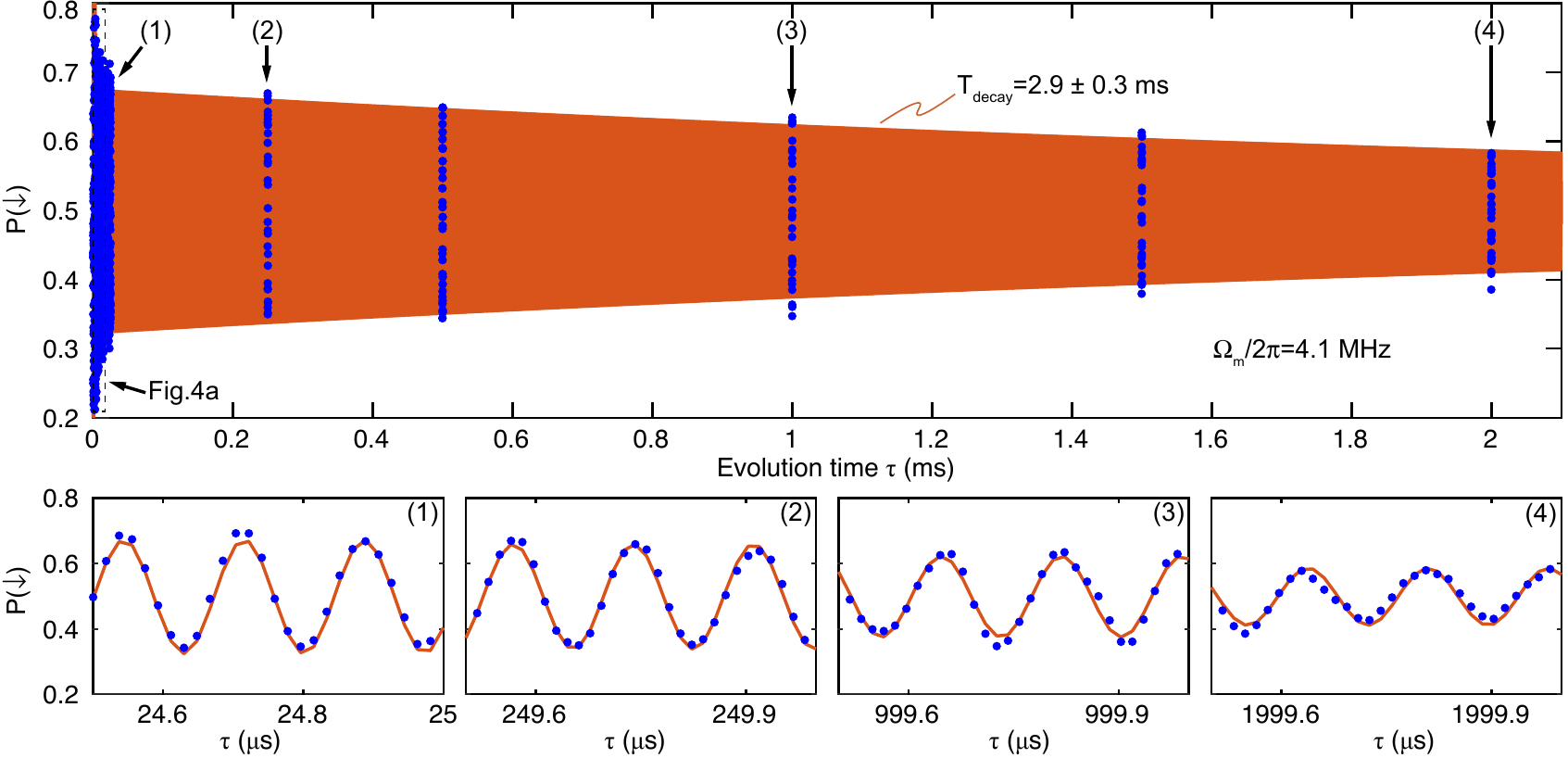} 
\caption{\label{FigLongRabi} 
a.) Single spin Rabi oscillations stabilised by parametric, mechanical driving with $\Omega_m=2\pi\times4.1$~MHz. Data (blue dots) are well-fit by an exponentially damped oscillation (orange) with decay time $T_{\rm decay}=2.9\pm0.3~$ms and a spin-precession (Rabi) frequency $5.83~$MHz. 
The quickly decaying transient at $\tau<20~\mu$s corresponds to the sidebands observed in Fig.\,\ref{FigMollow}a (see also Fig.\,\ref{FigCoherence}). The sub-panels show zoomed views over the measurement intervals labelled $(1)-(4)$ and confirm that the slowly decaying signal indeed consists of monochromatic Rabi oscillations at the mechanical frequency $\omega_m/2\pi$.
}
\end{figure*}

This phenomenon of frequency-locking is at the heart of our HCDD scheme and indeed efficiently decouples the NV spin from environmental fluctuations. 
The parametric drive couples the microwave dressed states  $\ket{\pm_N}$ and thereby yields new eigenstates $\ket{\pm_{N,M}}$, now doubly dressed by $N$ photons and $M$ cantilever phonons\,\cite{Rohr2014}.
For resonant strain-driving ($\Omega_{\rm MW}=\omega_m$), $\ket{\pm_{N,M}}=(\ket{-_N,M+1}\pm\ket{+_N,M})/\sqrt2$, where $M$ denotes the mean number of phonons in the cantilever, and $\ket{+_{N,M}}$ is split from $\ket{-_{N,M}}$ by an energy $\hbar\Omega_m$.
The resulting, doubly-dressed energy spectrum is illustrated in Fig.\,\ref{FigMollow}b as a function of $\Omega_m$, along with the possible transitions allowed between adjacent dressed states. These transitions are indeed also observed in the experimentally measured spin-precession spectra (coloured dashed lines in Fig.\,\ref{FigMollow}a).
The transition with the largest spectral weight, $\ket{\pm_{N,M}}\leftrightarrow \ket{\pm_{N,M-1}}$, occurs at $\omega_{\rm m}/2\pi$ and corresponds to a transition that changes the phonon number $M$ at constant microwave photon number $N$. 
The data presented in Fig.\,\ref{FigMollow}a indicate how doubly dressing using mechanically induced parametric strain driving protects the Rabi oscillations from environmental noise and prolongs their decay time: All involved energy levels  are insensitive to first-order to fluctuations of $\Omega_{\rm MW}$ around $\omega_m$, with the central peak at a precession frequency $\omega_m/2\pi$ insensitive to arbitrary orders. The same holds for vulnerability to microwave detunings (i.e. $\omega_{\rm MW}\neq\omega_0$); the only perturbation which affects the energies of the doubly dressed states to first order are fluctuations in $\Omega_m$, which are intrinsically low (see below).

This coherence protection through double dressing is already visible in the width of the dominant central frequency component (Fig.\,\ref{FigMollow}a), which is significantly narrower than all other spectral features but still limited by the measurement bandwidth $1/\tau_{\rm max.}$. To determine the intrinsic linewidth of this spectral feature, we conducted long-time Rabi oscillation measurements for resonant driving at $\Omega_{\rm MW}=\omega_{\rm m}$. The result (Fig.\,\ref{FigLongRabi}) shows sustained, coherent Rabi oscillations at frequency $\omega_{\rm m}/2\pi$ with a characteristic exponential decay over $T_{\rm Rabi}=2.9\pm0.3$~ms. This value is close to three orders of magnitude longer than the Gaussian decay time determined earlier without parametric mechanical driving, and the exponential decay we find 
indicates that 
the decay being induced by rapidly fluctuating noise sources, i.e. not by microwave power fluctuations anymore.
 
 \begin{figure}[t!]
\includegraphics[scale=1]{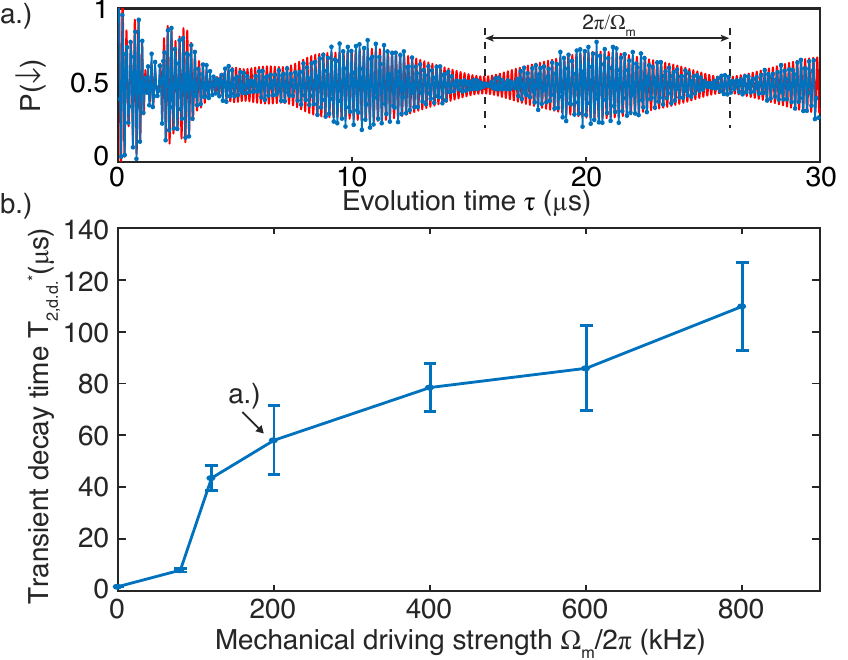} 
\caption{\label{FigCoherence} 
a.) Decay of transient Rabi oscillations for resonant parametric driving ($\Omega_{\rm MW}=\omega_m$) and $\Omega_m\approx2\pi\times200$~kHz. The transient is dominated by a beat-note at $\Omega_m$, i.e. the transient generates the Mollow-triplet sidebands visible in Fig.\,\ref{FigMollow}. The decay time of the beat-note (or, equivalently, the sideband linewidth) is a measure of the coherence time $T_{2,{\rm d.d.}}^*$ of the doubly-dressed spin states\,\cite{Cai2012a}. 
Data for $\tau\lesssim5~\mu$s also contain contributions from the off-resonantly driven, hsperfine-split NV spin transitions and are disregarded from the fit. 
b.) Coherence time $T_{2,{\rm d.d.}}^*$ as a function of mechanical driving strength $\Omega_m$, demonstrating increased coherence protection for increased mechanical driving. The observed coherence time saturates around $T_{2,{\rm d.d.}}^*\sim100~\mu$s due to technical limitations in our experiment (see text).
}
\end{figure}

The Rabi oscillation under parametric driving we observed allowed us to directly assess the coherence time $T_{2,{\rm d.d.}}^*$ of the doubly dressed spin states $\ket{\pm_{N,M}}$. 
As indicated in Fig.\,\ref{FigMollow}b, $\ket{+_{N,M}}$ and $\ket{-_{N,M}}$ are split in energy by $\Omega_m$, whose fluctuations of variance $\Delta\Omega_m$ thus directly set the inhomogenous dephasing time of the two-level system formed by $\ket{\pm_{N,M}}$ as $T_{2,{\rm d.d.}}^*=1/\sqrt{2}\pi\Delta\Omega_m$\,\cite{Jamonneau2016}.  
To assess $\Delta\Omega_m$ and therefore $T_{2,{\rm d.d.}}^*$, we determined the decay-time of the transient oscillations of the observed Rabi oscillations\,\cite{Cai2012a}, i.e. the width of the Mollow-triplet sidebands, which are mutually split by $\Omega_m$ (Fig.\,\ref{FigMollow}a).
Figure\,\ref{FigCoherence}a shows how these transient oscillations decay for $\Omega_m=2\pi\times200~$kHz, with a Gaussian decay-time $\tau_{\rm decay}=T_{2,{\rm d.d.}}^*=59~\mu$s. 
The coherence time is monotonically increasing with mechanical driving strength $\Omega_m$ (Fig.\,\ref{FigCoherence}b), as expected for concatenated decoupling by continuous driving. 
This increase, however, does not persist beyond $\Omega_m=2\pi\times800~$kHz (for which $T_{\rm d.d.}^*=110\pm17~\mu$s), due to 
increased mechanical noise (presumably due to nonlinearities of our diamond oscillator) at these high driving amplitudes. 
This deterioration is already visible in Fig.\,\ref{FigLongRabi}, where $\Omega_m=2\pi\times4.1$~MHz, and the initial, transient amplitude oscillations decay on a fast timescale $T_{\rm d.d.}^*\approx4~\mu$s. Extending $T_{\rm d.d.}^*$ further by increasing $\Omega_m$ would be possible through mechanical oscillators which yield higher strain fields, while avoiding nonlinearities when driven at high amplitudes\,\cite{MacQuarrie2015}. 
  
Our novel continuous decoupling scheme takes advantage of double-dressing with photons and phonons to enhance spin coherence of NV spins. 
In that sense, it bears strong similarity to the recently demonstrated decoupling by concatenated driving\,\cite{Cai2012a}.
Here, however, we also take advantage of the properties of our diamond mechanical oscillator for amplitude noise filtering, which in principle eliminates the need for further, higher-order decoupling fields. Indeed, mechanical resonators in general act as a low pass filters for amplitude-noise, with a cut-off frequency set by the mechanical linedwidth $f_c=\omega_m/(2\pi Q)$, with $Q$ the mechanical quality-factor. For our experiment under ambient conditions, we find $Q\approx530$, 
and therefore $f_c\approx11$~kHz, which still poses an impoprtant limitation to the coherence protection we can achieve. 
We note that under vacuum conditions, $Q\gtrsim10^6$ was reported for diamond mechanical oscillators\,\cite{Tao2014}, which would then yield $f_c\approx5$~Hz$\ll1/T_1$ and decoupling from the environment and driving field noise up to the ultimate limit imposed by the spin lifetime, $T_1$\,\cite{Cai2012a}.
  
To conclude, we have demonstrated a novel hybrid continuous dynamical decoupling scheme for a single spin that combines resonant microwave excitation with parametric driving of the spin by using strain generated in a nanomechanical oscillator. 
With this approach, we decoupled the spin from environmental noise and 
extended both the coherence time (from the typical $T_2^*\approx2~\mu$s to $T_{\rm d.d.}^*\gtrsim100~\mu$s) and the Rabi decay time (from $5.3~\mu$s to $2.9~$ms). 
Next experimental steps may include the use of high quality-factor mechanical oscillators for coherence protection up to the $T_1$-limit and the demonstration of coherent manipulation of dressed spin states.
Our work  thereby offers attractive perspectives for employing hybrid continuous dynamical decoupling of NV center spins for applications in quantum information processing and quantum sensing.

We thank A. Retzker, M. Kasperczyk, M. Munsch, B. Shields and E. Oudot for fruitful discussions and valuable input. We gratefully acknowledge financial support through the NCCR QSIT, a competence center funded by the Swiss NSF, through the Swiss Nanoscience Institute, 
by the EU FP7 project DIADEMS (grant \#611143)
and through SNF Project Grant 200021\_143697/1.

\bibliographystyle{apsrev4-1}
\bibliography{BibMollow}








\end{document}